\documentclass[12pt,preprint]{aastex}

\lefthead{Skinner et al.}
\righthead{Westerlund 1}

\begin{document}

\title{The {\em Chandra} X-ray Spectrum of the 10.6 s Pulsar \\
       in Westerlund 1: Testing the Magnetar Hypothesis}

\author{S.L. Skinner}
\affil{CASA, Univ. of Colorado, Boulder, CO 80309-0389 }

\author{R. Perna}
\affil{JILA, Univ. of Colorado, Boulder, CO 80309-0440}

\author{S.A. Zhekov}
\affil{Space Research Institute, Moskovska str. 6, Sofia-1000, Bulgaria}

%
\newcommand{\ltsimeq}{\raisebox{-0.6ex}{$\,\stackrel{\raisebox{-.2ex}%
{$\textstyle<$}}{\sim}\,$}}
%
\newcommand{\gtsimeq}{\raisebox{-0.6ex}{$\,\stackrel{\raisebox{-.2ex}%
{$\textstyle>$}}{\sim}\,$}}

\begin{abstract}
Two sensitive {\em Chandra} X-ray observations of the heavily-reddened
galactic starburst cluster Westerlund 1 in May and  June 2005 detected a 
previously unknown X-ray pulsar (CXO J164710.20$-$455217). Its slow
10.6 s pulsations, moderate X-ray temperature kT $\approx$ 0.5 keV, and
apparent lack of a massive companion tentatively suggest that it is an 
Anomalous X-ray Pulsar (AXP). An isothermal blackbody model
yields an acceptable spectral fit but the inferred source radius
is much less than that of a neutron star, a result that has also
been found for other AXPs. We analyze the X-ray spectra with more
complex  models including a model that assumes the pulsar is a
strongly magnetized neutron star (``magnetar'') with a light
element atmosphere.  We conclude that
the observed X-ray emission cannot be explained as global surface
emission arising from the surface of a cooling neutron star or
magnetar. The emission likely arises in one or more localized
regions (``hot spots'') covering
a small fraction of the surface. We discuss these new results in the 
context of both accretion and magnetar interpretations for the
X-ray emission.
\end{abstract}


\keywords{open clusters and associations: individual (Westerlund 1) ---
          stars: formation ---  stars: neutron ---
          X-rays: stars}

%

\newpage
\clearpage
\section{Introduction}
X-ray pulsars most commonly occur as members of binary
systems and their primary energy source is believed to be 
accretion from a donor star onto the  neutron star.
However, about a half dozen X-ray pulsars have been identified over 
the past decade that do not show the telltale Doppler shifts
associated with orbital motion and have slow
pulsation periods of $\approx$ 6 - 12 s and similar X-ray 
properties.  They are now known collectively as
Anomalous X-ray Pulsars (AXPs). The apparent absence of
a donor star, slow pulsation periods, and X-ray luminosity
levels  of AXPs suggest that their X-ray emission  is not 
powered by binary accretion or rotational energy loss. 
Two somewhat different pictures have emerged to explain
the X-ray emission. It has been proposed that 
AXPs are single neutron stars with intense magnetic fields 
B $\gtsimeq$ 10$^{14}$ G known as ``magnetars'' 
(Duncan \& Thompson 1992)  whose energy release
via  magnetic field decay heats the neutron star surface
to X-ray emitting temperatures (Thompson \& Duncan 1996 = TD96).
This idea is attractive but a direct measurement to confirm
the presence of the very high magnetic field strengths postulated for
magnetars is still needed. A second possibility is that the
X-ray emission may be due to accretion from a residual disk
(van Paradijs et al. 1995; Chatterjee, Hernquist, \& Narayan 2000;
Alpar 2001).

The X-ray properties  of AXPs have been summarized by 
Mereghetti et al. (2002 = M02) and Kaspi \& Gavriil 
(2004 = KG04). AXPs have moderately soft X-ray 
spectra and X-ray luminosities in the range
L$_{\rm X}$ $\sim$ 10$^{34}$ -  10$^{35}$ ergs s$^{-1}$.
Acceptable spectral fits generally require two components.
In most cases the emission is modeled with a thermal
(e.g. blackbody) component at kT$_{bb}$ $\approx$ 0.4 - 0.6 keV
plus a harder power-law
component. A wide range of photon power-law indices
$\alpha_{ph}$ $\approx$ 2 - 5 has been reported 
in the literature (M02, Perna et al. 2001 = P01).
The origin of the non-thermal power-law emission is  
uncertain but it is thought to arise in the 
magnetosphere (M02). 

The interpretation of the cool thermal X-ray emission as
integrated blackbody emission from the surface
of the neutron star is problematic for AXPs because the
source radius inferred from isothermal
blackbody fits is usually only a small fraction of the 
radius of a neutron star (R$_{ns}$ $\approx$ 10 km).
This result  could be explained if the intrinsic
surface emission is masked by more luminous X-ray
emission originating in hot spots that cover only a 
small fraction of the stellar surface. 
Gotthelf \& Halpern (2005, = GH05) have suggested that this
is the case for the emission of the transient 
AXP XTE J1810$-$197 monitored during the decay phase of a recent
outburst. They model its  emission as a concentric
spot arrangement using a two-temperature blackbody
model and argue that this model is more physically
meaningful for XTE J1810$-$197 than the usual 
blackbody $+$ power-law interpretation.

The spot interpretation offers a promising means of explaining
the variable X-ray emission of transient AXPs such as XTE J1810$-$197
but the origin of quiescent emission from AXPs is another 
question. The issue of  small source radii inferred from  
blackbody models of quiescent AXPs was explored by P01.
They noted that the intrinsic X-ray spectrum
of a magnetar will differ from that of a simple
blackbody due to the effects of anisotropic heat flow
through the envelope, reprocessing of photons
by a light-element atmosphere, and general 
relativistic corrections. Thus, source radii inferred
from blackbody fits of magnetars may be incorrect. They developed
a sophisticated magnetar model incorporating the
above effects and used it to fit the  {\em ASCA} X-ray
spectra of five AXPs. The source radii inferred from
their magnetar models were generally consistent with 
the radius of a neutron star, but interestingly their
spectral fits still required a power-law component.
The X-ray temperatures determined from their magnetar
models were kT$_{magnetar}$ $\approx$ 0.3 - 0.4 keV,
which are slightly lower than the blackbody temperatures
but still higher than expected for thermal cooling alone.

Since only a handful  of  AXPs are presently known, additional
objects are needed to fully characterize their X-ray properties
and guide the development of realistic X-ray emission models.
We analyze here the X-ray spectrum of
a new 10.6 s pulsar that was  serendipitously detected in two
sensitive {\em Chandra} observations of the galactic starburst
cluster  Westerlund 1 (Wd1) in May - June 2005 
(Skinner et al. 2005 = S05; Muno et al. 2006 = M06;
Skinner et al. 2006 = S06). This object shows several of 
the characteristic X-ray properties of AXPs including 
slow pulsations, a thermal component temperature
kT $\approx$ 0.5 - 0.6 keV, a possible power-law contribution, 
and an inferred emitting area from isothermal blackbody
fits that is  much less than the surface area of a 
neutron star. {\em NTT} images show no evidence
for an infrared counterpart down to a limiting magnitude
K$_{s}$ = 18.5 (S05, M06) which initially seems to 
rule out a massive ($>$1 M$_{\odot}$) companion
and thus strengthens the AXP classification. Even so,
thermal emission models indicate that the Wd 1 pulsar
is less luminous in X-rays than other AXPs based on
current distance estimates.

It has recently been argued that the 10.6 s pulsar
in Wd 1 is a magnetar (M06). However, further observational
work will be needed to confirm this hypothesis.
The existing constraint on the spin-down rate
from two {\em Chandra} observations spaced $\approx$1 month
apart is not yet sufficient to rule out spin-down as the X-ray
energy source. In addition, more sensitive optical/IR/millimeter 
searches are necessary to determine if a low-mass companion
or residual disk are present. It is our objective here
to examine more sophisticated X-ray spectral models 
for the Wd 1 pulsar that go beyond the isothermal blackbody
model considered in earlier work. Most importantly, we
consider a model that assumes the neutron star is a magnetar
and show that the source radius inferred from the magnetar 
model is still much less than that  of a neutron star.
This result gives support to the idea that the detected
X-ray emission originates mainly in one or more localized
regions (hot spots) covering only  a small fraction of
the neutron star surface. Such  spots could form on 
the surface of a magnetar but might also occur as
heated polar caps on a conventional X-ray pulsar.

\section{Chandra Observations and Data Reduction}

{\em Chandra} observed Westerlund 1 with the ACIS-S imaging array
on 22-23 May 2005 and 18-19 June 2005  with exposure
live times of 18,808 s and 38,473 s, respectively. 
Additional observational details are given in S06.
The observation
was obtained in timed faint-event mode using a 3.2 s frame time.
For the spectral analysis discussed below, Level 1 event files
provided by the {\em Chandra} X-ray Center were 
processed with CIAO 
\footnote{Further information on {\em Chandra} Interactive
Analysis of Observations (CIAO) software can be found at
http://asc.harvard.edu/ciao.} (vers. 3.3) to generate an
updated Level 2 event file that takes into account the latest
gain and effective area calibrations (CALDB vers. 3.2).
Spectra and associated  instrument response files were then
extracted for each observation using the 
CIAO tool {\em psextract} using a circular
extraction region of radius R$_{ext}$ = 2.$''$5 centered on the X-ray source.
Background was extracted from adjacent source-free regions
on the same CCD and is negligible, comprising only about
0.7\% of the total counts within the source extraction region.
Spectra were rebinned to a minimum of 20 counts per bin and
analyzed using XSPEC vers. 12.2.0bp.
We integrated the magnetar atmosphere model described 
by P01 into XSPEC for spectral fitting.

\section{Pulsar X-ray Properties}

The pulsar was identified on the basis of 10.61 s pulsations 
discovered in the  brightest X-ray source detected in the
{\em Chandra} Wd 1 observations, lying  $\approx$1.6$'$
southeast of the Wd 1 core (Fig. 1 of S06; Fig. 2 of M06).
Timing analysis was discussed by M06 and will not be 
discussed further in this paper. Higher time resolution
observations will be needed to characterize the pulse properties.

The {\em Chandra} position of the pulsar as determined by
the CIAO {\em wavdetect} wavelet detection tool using data
from the longer exposure on 18-19 June 2005 is
(J2000.0) R.A. $=$  16$^h$ 47$^m$ 10.$^s$20, 
decl. $=$ $-$45$^{\circ}$ 52$'$ 17.05$''$. The 90\% source 
location error circle for ACIS-S 
\footnote{Chandra Proposer's Observatory Guide, Rev. 8.0,
Dec. 2005; http://cxc.harvard.edu }
has a radius of $\approx$0.$''$5. The X-ray
centroid obtained from the {\em XIMAGE} image analysis 
software package is nearly identical.
The May 2005 observation detected 386 $\pm$ 20 net counts 
and the June observation yielded 834 $\pm$ 29 net counts,
based on events in the 0.3 - 7 keV range. 
The respective count rates were 
2.05 ($\pm$ 0.11) $\times$ 10$^{-2}$  c s$^{-1}$
and 2.17 ($\pm$ 0.08)  $\times$ 10$^{-2}$  c s$^{-1}$,
which are the same to within the uncertainties. Photon pileup
is negligible ($\leq$3\%). 
Since no optical or IR counterpart has yet been found, 
cluster membership has  not been proven and the 
distance is uncertain. Spectroscopic studies give a
distance to Wd 1 in the range d = 2 - 5.5 kpc
(Clark et al. 2005 = C05) and the photometric study 
of Brandner et al. (2005 = B05) gives d = 4 kpc.
In the following we adopt a distance d = 5 kpc to
the pulsar based on the assumption that it is associated
with Wd 1.

Figures 1 and 2 show the ACIS-S spectrum of the pulsar. 
The spectrum
is heavily absorbed below $\approx$1 keV. Strong absorption is 
anticipated if the pulsar lies in (or behind) Wd 1 based on cluster 
extinction estimates and  the total galactic HI column density toward 
the pulsar, which is N$_{\rm H}$ = 2.2 $\times$ 10$^{22}$ cm$^{-2}$
based on the HEASARC N$_{\rm H}$ calculation tool
\footnote{http://heasarc.gsfc.nasa.gov/cgi-bin/Tools/w3nh/w3nh.pl}.
Cluster extinction estimates are in the range
A$_{\rm V}$ $\approx$ 9.5 - 13.6 mag (B05; C05) which 
corresponds to N$_{\rm H}$ $\approx$ (2.1 - 3.0) 
$\times$ 10$^{22}$ cm$^{-2}$ using the conversion of
Gorenstein (1975). Our spectral fits of the Wd 1 pulsar
with models that include a thermal (e.g. blackbody) component
yield N$_{\rm H}$ estimates that are
at or slightly below the low end of this range (Table 1).

The pulsar spectrum shows no obvious lines with the possible 
exception of a weak emission feature near 3.5 keV. As Figure 3
shows, this feature is  most apparent in the longer second
observation but is only weakly present (if at all) in the
shorter first observation. Since the feature is not clearly
present in the first observation, its reality is in doubt.
There are no known thermal emission lines near kT = 3.5 keV 
or near kT =  4.5 keV, where the latter value is corrected for
gravitational redshift (M$_{ns}$ = 1.4 M$_{\odot}$, R$_{ns}$ =
10 km). We are not aware of any instrumental effects that 
would give rise to a feature near 3.5 keV. There is a gold M
edge at 3.428 keV but a detectable {\em emission} feature from
this edge seems unlikely. If the feature is confirmed as an
emission line in higher signal-to-noise spectra, a nonthermal 
interpretation will likely be required.

\section{X-ray Spectral Models}

For the spectral analysis discussed below, we focus on the
spectrum from the second observation in June 2005, which provides
the best signal-to-noise ratio. We attempted to fit the spectrum with 
a variety of emission models, as summarized in Table 1. 
All models included an absorption component based on cross-sections 
from Morrison \& McCammon (1983).

\subsection{Power Law Model}

Before considering models that include a thermal 
emission component, 
we note for  completeness that the spectrum can
be acceptably fitted with a simple power-law model (model A in Table 1).
The N$_{\rm H}$ and unabsorbed X-ray luminosity 
log L$_{X}$ = 34.44 ergs s$^{-1}$ determined from this power-law 
model are larger than obtained from models that include
thermal emission and the inferred  N$_{\rm H}$ is at or above 
the maximum expected from A$_{\rm V}$ estimates (Sec. 3).
Simple power-law models have been  used to fit the soft-band
X-ray spectra of some rotation-powered pulsars (Becker \&
Tr\"{u}mper 1997), but acceptable fits of AXPs based on
good-quality X-ray spectra  typically require at least two emission 
components (e.g. blackbody $+$ power-law). Thus, if the 
Wd 1 pulsar is an AXP then the ability to fit its spectrum
with a simple power-law model may be a consequence of
limited signal-to-noise in the existing {\em Chandra} data.
The  rather high  N$_{\rm H}$ deduced from the power-law
model motivates us to consider models that include a 
thermal component, as discussed below.

\subsection{Blackbody Models}

An isothermal blackbody model (model B in Table 1) yields an
acceptable fit with kT$_{bb}$ = 0.59 keV. 
Although the fit is formally acceptable ($\chi^2$/dof = 34.7/35 = 0.99),
X-ray modeling of other AXPs (P01) suggests that a power-law
component may also be present. A two-component blackbody $+$
power-law model (model C in Table 1) provides only a minor
improvement in the $\chi^2$ fit statistic. Thus, the existing
{\em Chandra} data are consistent with a two-component 
blackbody $+$ power-law model but it is not necessary to
include both components to obtain an acceptable fit.
Our best-fit  blackbody $+$ power-law model gives a
photon power-law index $\alpha_{ph}$ $\approx$ 1.8, but this value
is not tightly constrained. In this  blackbody $+$ power-law model
the power-law component could contribute as much as 
$\approx$40\% of the observed
(absorbed) flux in the 0.3 - 8 keV range (Table 1). 

The inferred source radius is in the range R$_{s}$ = 0.27 -
0.36 km  for the two blackbody models in Table 1. These
values are much less than the radius of a neutron star.
It is thus obvious that we are not detecting thermal
emission emanating from the surface of a neutron star,
assuming the intrinsic spectrum is an isothermal blackbody.
As already noted (Sec. 1), blackbody models may yield
incorrect radii due to atmospheric effects. We investigate
this possibility below in the case where the X-ray
source is assumed to be a magnetar.

\subsection{Magnetar Models}

The magnetar model developed by P01 assumes an underlying
highly magnetized neutron star of mass M$_{ns}$ = 1.4 M$_{\odot}$ 
cooling via blackbody emission  that is reprocessed through
a light-element highly magnetized atmosphere (Heyl \& Hernquist 1998).
The model computes the phase-averaged flux as a function
of photon energy $E$, taking into account anisotropic heat flow 
through the envelope due to magnetic field effects 
and general relativistic light deflection
(eqs. 3-5 of P01).  The phase-averaged
magnetar flux  $F(E)$ at a distance
$d$ from the star to observer without any correction for
absorption by intervening material  is (eq. [3] of P01 in
corrected form):

\begin{equation}
F(E) = \frac{R_{\infty}^2 \sigma T_{p,\infty}^4}{d^2} 
       \frac{1}{kT_{p,\infty}}
       \times
       \int_{0}^{2\pi}{\frac{d\alpha}{2\pi}} 
       \int_{0}^{1}{2x~dx} 
       \int_{0}^{2\pi}\frac{d\phi}{2\pi}I_{0}(\theta,\phi)n(Ee^{-\xi},T_{s}(\theta,\phi))
\end{equation}

In the above, $T_{p,\infty}$ $\equiv$ T$_{p}e^{\xi}$ where
T$_{p}$ is the pole temperature, 
$R_{\infty}$ $\equiv$ R$_{ns}e^{-\xi}$, and
$e^{\xi}$ $\equiv$  $\sqrt{1 - (R_{sch}/R_{ns})}$ where
$R_{ns}$ is the stellar radius, $R_{sch}$ = 2GM$_{ns}$/c$^{2}$
is the Schwarzschild radius,
and the integrated spectrum over the stellar surface accounts for 
general relativistic light deflection and the effects of a
light-element atmosphere.
The angle $\alpha$ is the angle between
the magnetic pole and line-of-sight, $\delta$ is the angle
between an emitted photon and the normal to the surface, and
$x$ $\equiv$ sin $\delta$. The spherical coordinate angles 
($\theta$,$\phi$) specify the position on the surface of
the star, $I_{0}(\theta,\phi)$ is the surface intensity  
distribution (eq. [5] of P01), and 
$n(Ee^{-\xi},T_{s}(\theta,\phi)$) specifies the 
local emission at a point on the surface as a function
of the local temperature $T_{s}$ (eq. [6 ] of P01).

We incorporated the magnetar model of P01 into XSPEC
and reran the blackbody fits discussed above, replacing
the blackbody model with the magnetar model. The magnetar
fit results are summarized as models D and E in Table 1.
In our implementation, the radius of the 
neutron star and the stellar distance were specified
as input parameters. 
For the fits in Table 1 we assumed R$_{ns}$ = 10 km and
d = 5 kpc. The model was used to find the
best-fit value of the neutron star pole temperature
and flux normalization factor 
norm$_{1}$ = F$_{X,th}$/F$_{X,mag}$, where F$_{X,th}$ 
is the unabsorbed flux due to the thermal component 
over the fitted energy range and  F$_{X,mag}$ is the flux 
predicted by the model for a magnetar of the assumed radius 
and distance.

As Table 1 shows, the $\chi^2$ fit statistics for the magnetar
models are comparable to or slightly better than those
for the blackbody models. The overall fit for the two-component
magnetar $+$ power-law  model (model E) shown in Figure 4 looks
nearly identical to that of the blackbody $+$ power-law model
(Fig. 1). However, a comparison of the unfolded spectra in
Figures 2 and 5 shows that contribution of a power-law component
is much less in the magnetar $+$ power-law  model. The overall
shape of the spectrum is matched quite well with the magnetar
component alone, and any power-law component need not contribute
more than  $\approx$15\%  of the observed flux (0.3 - 8 keV).

The inferred magnetar pole 
temperatures are  in the range kT$_{p}$ = 0.44 - 0.48 keV. 
These values are
10\% - 20\% less than those obtained with the corresponding
blackbody models, but still too high to be reconciled with thermal
cooling alone. A similar result was noted by P01 when fitting
{\em ASCA} spectra for five AXPs, and they concluded  that 
additional heating (possibly by magnetic field decay) is
needed to explain the high temperatures. 

It is apparent from Table 1 that the flux normalization factor
norm$_{1}$ is much less than unity for the magnetar 
models. By definition of  norm$_{1}$ (see above), this
indicates that the unabsorbed flux predicted by the magnetar model 
with  R$_{ns}$ = 10 km and d = 5 kpc is   $\sim$10$^{2}$ greater 
than that determined from the X-ray spectrum. Since the leading
term in the magnetar flux relation (eq. 1) is the  blackbody flux,
the flux mismatch indicates that the source radius is much
less than  R$_{ns}$, provided that d = 5 kpc is not a serious
underestimate.  This is shown in the last row of Table 1,
which gives the source radius R$_{s}$ determined from the 
blackbody formula (which we
emphasize is only an approximation in the magnetar case).
The discrepancy could be removed by assuming
the pulsar is $\sim$10 times more distant, but this seems unlikely
if the pulsar is  indeed associated with Wd 1.
We thus conclude that the observed X-ray emission 
cannot be readily explained as global emission coming
from the surface of a magnetar cooling through a 
light-element atmosphere.

\section{Discussion}

The spectral analysis in Section 4  shows the following:
(i) most of the observed X-ray flux in the {\em Chandra} bandpass
can be accounted for by
thermal models but a possible power-law contribution
is not ruled out by the existing data,
(ii) thermal models give  a characteristic temperature 
in the range kT $\approx$ 0.4 - 0.6 keV, which is higher
than expected for neutron star cooling alone, 
(iii) the inferred source radius from blackbody models is
R$_{s}$ $<$ 0.4 km and (iv) a magnetar model for a neutron
star of radius R$_{ns}$ = 10 km emitting from its entire 
surface overestimates the flux, indicative of 
an emitting region that is considerably  smaller than R$_{ns}$
unless the pulsar distance d = 5 kpc is significantly underestimated.

Thus, the models considered here point to a small emitting
region of relatively warm plasma (T $\sim$ 5 - 7 MK) as the origin  
of most of the observed X-ray emission. A plausible explanation
is that the emission arises from one or more high-temperature
regions or ``hot spots'' on the neutron star surface, rather than 
from the entire surface of a cooling neutron star. 
We obtain an equivalent blackbody source radius in the 
range  R$_{s}$ =  0.27 - 0.52 km (Table 1), but this does
not necessarily correspond to the radius of any particular
spot since multiple spots may be present. We have assumed
d = 5 kpc but the conclusion of a small emitting
region R$_{s}$ $<<$ R$_{ns}$ holds even if the distance is 
twice that value. The ratio of source emitting area to
stellar surface area (R$_{s}$/R$_{ns}$)$^{2}$ 
$\sim$ 10$^{-3}$ obtained here is comparable to that obtained
for some  field pulsars with magnetic field strengths below 
the magnetar range (e.g. Greiveldlinger et al. 1996)
but is smaller than the typical ratio 
(R$_{s}$/R$_{ns}$)$^{2}$ $\sim$ 10$^{-1}$ derived for the 
AXPs studied by Durant \& van Kerkwijk (2006).

Spots can form on a magnetar as a result of the  local concentration 
of magnetic field lines (TD96). In that case, the equivalent blackbody
temperature of a spot is expected to anti-correlate with the spot radius
for objects having the same L$_{X}$ (eq. [92] of TD96). The calculation
of  equivalent  blackbody radii made possible by new distance determinations 
of several galactic AXPs (Durant \& van Kerkwijk 2006) suggests that
such an  anti-correlation may be present.

Spots can also form as a result of accretion onto the neutron star,
even if it is not a magnetar.
Assuming that the Wd 1 pulsar does not have a low-mass
companion, the accretion reservoir could in principle be a 
fallback disk created after the supernova explosion
(Chatterjee et al. 2000; Alpar 2001)
or the ISM (Blaes \& Madau 1993). More sensitive images in the infrared
and at millimeter wavelengths are needed to determine if a disk is
present. This question is of considerable interest given the recent
detection of mid-IR emission at the position of the isolated young pulsar
4U 0142$+$61 (Wang et al. 2006). These authors argue that the 
mid-IR emission arises in a passive X-ray heated disk around the
neutron star that may have originated from supernova ejecta that 
subsequently fell back onto the neutron star.

Until observational evidence for a disk around the Wd 1 pulsar is
presented, any conclusions based on disk models should be considered
speculative. We only remark that if the pulsation period 
P = 10.6 s is close to the 
equilibrium period for uniform spin-down by a disk (eq. [6] of
TD96) then a dipole field strength  B$_{dipole}$
$\sim$ 10$^{11}$ G. is inferred from L$_{X}$ (Table 1).
This field strength is a lower limit if the detected emission
is dominated by spots since the total L$_{X}$ would contain
an additional contribution from the cooling surface of the
neutron star that may be masked by absorption.
Accretion from the ISM onto polar caps could achieve 
temperatures  kT$_{bb}$ $\approx$ 0.5 keV (Table 1) for
realistic accretion rates (eq. [29] of Blaes \& Madau 1993). 
Even so,  the accretion rate and gas density around the
source needed to account for  L$_{X}$ $\sim$ 10$^{33}$ ergs s$^{-1}$ 
(Table 1)  are uncomfortably high and the ISM accretion picture
is difficult to justify.

If the observed emission is coming predominantly from one or more
hot spots, then where is the global emission from the neutron star
surface? A likely explanation is that  the surface
emission is largely absorbed due to the relatively  high intervening
absorption column toward the pulsar. If we assume a soft intrinsic
blackbody spectrum with kT$_{bb}$ = 0.1 keV for a cooling neutron 
star with R$_{ns}$ = 10 km  then the unabsorbed broadband flux  
at d = 5 kpc is 
F$_{X}$ = 4.3 $\times$ 10$^{-13}$ ergs cm$^{-2}$ s$^{-1}$.
Assuming N$_{\rm H}$ = 
1.7 $\times$ 10$^{22}$ cm$^{-2}$ (a typical value from Table 1),
the {\em PIMMS}
\footnote{http://asc.harvard.edu/toolkit/pimms.jsp}
simulator predicts that no more that $\approx$5
counts would have been detected in the second {\em Chandra} 
ACIS-S exposure (38.5 ksec) from this hypothetical blackbody.
If the neutron star is a magnetar, then theoretical cooling curves
(Heyl \& Kulkarni 1998) predict kT $\leq$ 0.1 keV for ages
$t$ $\gtsimeq$ 0.2 Myr (B $\sim$ 10$^{14}$ G).

Finally, we comment on the X-ray luminosity of the Wd 1 pulsar
and its relevance to the AXP classification and the energy
source that powers the X-ray emission. 
The maximum X-ray luminosity determined from models that 
include a thermal component is log L$_{X}$ = 33.20 ergs s$^{-1}$
(model C in Table 1).  This L$_{X}$ along with the inferred
blackbody temperature kT$_{bb}$ $\approx$ 0.5 keV places the Wd 1 
pulsar in the (kT$_{bb}$,L$_{X}$) plane at a temperature similar
to other known AXPs but at a luminosity that is at least $\sim$10 times
lower (Fig. 5 of  M02). This conclusion is strengthened by the recent
work of Durant \& van Kerkwijk (2006), who conclude that AXP X-ray
luminosities are typically L$_{X}$ $\sim$ 10$^{35}$ ergs s$^{-1}$.  
It thus appears that the 
Wd 1 pulsar is underluminous in X-rays relative to previously known AXPs. 

There are several possible explanations for the apparent low X-ray
luminosity,  apart from the obvious possibility of an underestimated 
distance. Since the
X-ray temperature obtained for the Wd 1 pulsar is similar to other
AXPs (P01) the lower L$_{X}$ may just be an indication of a smaller
emitting area (i.e. emission dominated by highly localized spots).  
Alternatively, if the Wd 1 pulsar and 
other AXPs are magnetars powered by magnetic
field decay, then the lower L$_{X}$ for the Wd 1 pulsar could be
an indication of lower magnetic
field strength (B). The surface heat flux of a magnetar
scales sensitively as B$^{4.4}$ (eq. [91] of TD96). Although  
L$_{X}$ for the Wd 1 pulsar does seem low, it should be 
kept in mind that only a few AXPs are presently known and
the AXP X-ray luminosity function is not well-sampled. 
Observational selection effects may have biased previous
AXP identifications toward more X-ray luminous members of the class.
Finally, the discrepancy could be resolved if the simple power-law
model (model A) is correct. However, as we have already noted,
a simple power-law model is likely an over-simplification of
the true intrinsic spectrum if the Wd 1 pulsar is an AXP.

Do we need to invoke the magnetar interpretation for
the Wd 1 pulsar? Could its X-ray luminosity 
be powered by spin-down instead of an ultra-strong
magnetic field? The {\em maximum} X-ray 
luminosity that can be derived from rotational energy release for a
pulsar of period P, age $t$, and moment of inertia $I$
is (eq. [2] of TD96)

\begin{equation}
L_X \sim  \frac{1}{2t}I\left[\frac{2\pi}{\rm{P}}\right]^2~\rm{ergs~s^{-1}} .
\end{equation}

\noindent Using P = 10.6 s and assuming M$_{ns}$ = 1.4 M$_{\odot}$ and
R$_{ns}$ = 10 km, the above becomes

\begin{equation}
t_{4} \sim \frac{6 \times 10^{32}}{ \rm{L_X}~(\rm {ergs~ s^{-1})}}
\end{equation}

\noindent where $t_4$ is the age in units of 10$^{4}$ yr. 
From the thermal-component models in Table 1
we have L$_{X}$ $\approx$ 1.4 $\times$ 10$^{33}$ ergs s$^{-1}$, but 
as noted above this L$_{X}$ should be considered a lower limit 
on the total X-ray luminosity (neutron star cooling surface $+$ hot spots).
In that case, spin-down is a plausible energy source
if the Wd 1 pulsar age is $t$ $\ltsimeq$4 $\times$ 10$^{3}$ yr.
The two {\em Chandra} observations give an upper limit on the 
period derivative \.{P} $<$ 2 $\times$ 10$^{-10}$ s s$^{-1}$
(M06). Using the relation
\.E = $-$4$\pi^2$$I$\.P/P$^{3}$ we obtain an
an upper limit on the rate of energy release
from spin-down of log \.E $<$ 33.87 ergs s$^{-1}$.
As can be seen (Table 1), the X-ray luminosities inferred from
thermal-component spectral fits are a few times less than the above value of \.E.



\section{Summary}

The X-ray emission detected by {\em Chandra} from the
10.6 s pulsar in Wd 1 is not global surface emission from a
cooling neutron star or magnetar. Both blackbody and magnetar
models imply an emitting region that is much smaller than
the radius of a neutron star. Models that include a 
thermal component show that the
observed emission most likely arises from
one or more hot spots (kT $\approx$ 0.5 keV) 
covering a small fraction of the surface.  
Any cool (kT $\leq$ 0.1 keV) emission from the cooling
surface would have been heavily attenuated by intervening
absorption.

The Wd 1 pulsar shows X-ray properties that justify its  
tentative classification as an AXP. Even so, the X-ray
luminosity L$_{X}$ $\approx$ 10$^{33.15}$(d/5 kpc)$^{2}$
ergs s$^{-1}$ (0.3 - 8 keV)  of  the  pulsar  deduced from  
thermal spectral models is at least an order of magnitude below
that of known AXPs if d $\approx$ 5 kpc. The existing
constraint on \.P from two {\em Chandra} observations is
not stringent enough to rule out spin-down as the energy
source if the pulsar is young and if its X-ray luminosity
is indeed as low as L$_{X}$ $\approx$ 10$^{33.15}$ ergs s$^{-1}$.

Since the pulsar has only recently been discovered, further 
observational work will be needed to clarify its nature.
Deeper optical/IR/millimeter searches for  a low-mass
companion or residual disk will be  particularly important for
the X-ray interpretation.  A  higher signal-to-noise
X-ray spectrum is needed to distinguish between competing
emission models and a tighter constraint on \.P from
continued X-ray time monitoring would be useful  to determine
if spin-down can account for the X-ray luminosity.

\acknowledgments

This research was supported by NASA  grant 
GO5-6009X.

\clearpage

%
%

\clearpage

\begin{deluxetable}{llllll}
\tabletypesize{\scriptsize}
\tablewidth{0pc}
\tablecaption{{\em Chandra} ACIS-S Spectral Fits for the Wd 1 Pulsar 
   \label{tbl-1}}
\tablehead{
\colhead{Parameter}      &
\colhead{ }        &
\colhead{  }
}
\startdata
Model\tablenotemark{a}              &  A                  &   B                 &         C             &       D            & E        \nl
Components                          &  pl                 & bb                  & bb$+$pl               & mag                & mag$+$pl   \nl
N$_{\rm H}$ (10$^{22}$ cm$^{-2}$)   & 3.1 [2.7 - 3.6]     & 1.5 [1.2 - 1.8]     & 1.8 [1.3 - 3.2]       & 1.7 [1.4 - 2.0]       & 1.8 [1.4 - 3.9] \nl
kT (keV)                            & ...                 & 0.59 [0.54 - 0.63]  & 0.50 [0.36 - 0.60]    & 0.48 [0.42 - 0.53] & 0.44 [0.31 - 0.52] \nl
norm$_{1}$\tablenotemark{b}         & ...                 & 0.30                & 0.53                  & 4.6e-03             & 6.1e-03  \nl 
$\alpha_{ph}$                       & 3.8 [3.4 - 4.2]     & ...                 & 1.8 [...]             & ...                  & 1.8 [...]   \nl
norm$_{pl}$                         & 1.2e-03             & ...                 & 3.1e-05               & ...                  & 1.2e-05     \nl
$\chi^2$/dof                        & 35.3/35             & 34.7/35             & 32.0/33               & 32.2/35            & 31.8/33      \nl
$\chi^2_{red}$                      & 1.0                 & 0.99                & 0.97                  & 0.92               & 0.96         \nl
F$_{\rm X}$\tablenotemark{c}
                                    & 2.14 (91.9)         & 1.93 (3.79)         & 2.18 (5.37)           & 2.00 (4.37)        & 2.09 (5.03)  \nl
F$_{\rm X,th}$\tablenotemark{c}
                                    & ...                 & 1.93 (3.79)         & 1.34 (3.56)           & 2.00 (4.37)        & 1.76 (4.32)  \nl
log L$_{\rm X}$ (ergs s$^{-1}$)     & 34.44               & 33.05               & 33.20                 & 33.12              & 33.18  \nl
R$_{s}$ (km)                        & ...                 & 0.27                & 0.36                  & 0.44               & 0.52   \nl
\enddata
\tablecomments{
Based on  XSPEC (vers. 12.2.0) fits of the background-subtracted ACIS-S spectrum of
the Wd 1 pulsar (CXO J164710.20$-$455217) binned 
to  20 counts per bin using 38,473 s ksec of  exposure obtained on
18-19 June 2005. Blackbody (bb) emission was modeled with the $bbodyrad$ model in XSPEC.
Magnetar (mag) emission was modeled with the custom model $magnetar$  (see text).
The tabulated parameters
are absorption column density (N$_{\rm H}$), blackbody temperature or magnetar pole temperature (kT),
normalization of the blackbody or magnetar component (norm$_{1}$), photon power-law index ($\alpha_{ph}$), 
power-law normalization (norm$_{pl}$).
Square brackets enclose 90\% confidence intervals and an ellipsis means that 
the algorithm used to compute confidence intervals did not converge.
The  total X-ray flux (F$_{\rm X}$) 
is  the absorbed value in the 0.3 - 8 keV range, followed in 
parentheses by  unabsorbed value. The thermal flux associated with the 
blackbody or magnetar component is F$_{\rm X,th}$.
The unabsorbed luminosity L$_{\rm X}$ (0.3 - 8 keV)  assumes a
distance of 5  kpc.
R$_{s}$ is the inferred blackbody source radius at d = 5 kpc
based on the relation R$_{s}^2$ = L$_{X,th}$/(4$\pi\sigma$T$^4$) where 
L$_{X,th}$ is the luminosity associated with the unabsorbed thermal flux F$_{X,th}$.
This relation is only approximate for models D and E since
magnetar spectra are not true blackbodies.
} 
\tablenotetext{a}{Model A:~N$_{\rm H}$$\cdot$(PL);
Model B:~N$_{\rm H}$$\cdot$(kT$_{bb}$);
Model C:~N$_{\rm H}$$\cdot$(kT$_{bb}$ $+$ PL);
Model D:~N$_{\rm H}$$\cdot$(kT$_{magnetar}$);
Model E:~N$_{\rm H}$$\cdot$(kT$_{magnetar}$ $+$ PL)
}

\tablenotetext{b}{For models B and C:~
                  norm$_{1}$ = R$^{2}_{km}$/d$_{10}^2$ where 
                  R$_{km}$ is the  source radius in km and 
                  d$_{10}$ is the distance to the source
                  in units of 10 kpc.  
                  For models D and E:~ 
                  norm$_{1}$ = F$_{X,th}$/F$_{mag}$ where 
                  F$_{X,th}$ is the unabsorbed thermal flux  flux 
                  determined from the spectrum and 
                  F$_{mag}$ is the theoretical flux predicted for
                  a magnetar with radius R = 10 km and distance d = 5 kpc
                  using the model of P01.}

\tablenotetext{c}{Flux units are 10$^{-13}$ ergs cm$^{-2}$ s$^{-1}$.}

\end{deluxetable}

\clearpage



\begin{figure}
\figurenum{1}
\epsscale{1.0}
\includegraphics*[width=12.5cm,angle=-90]{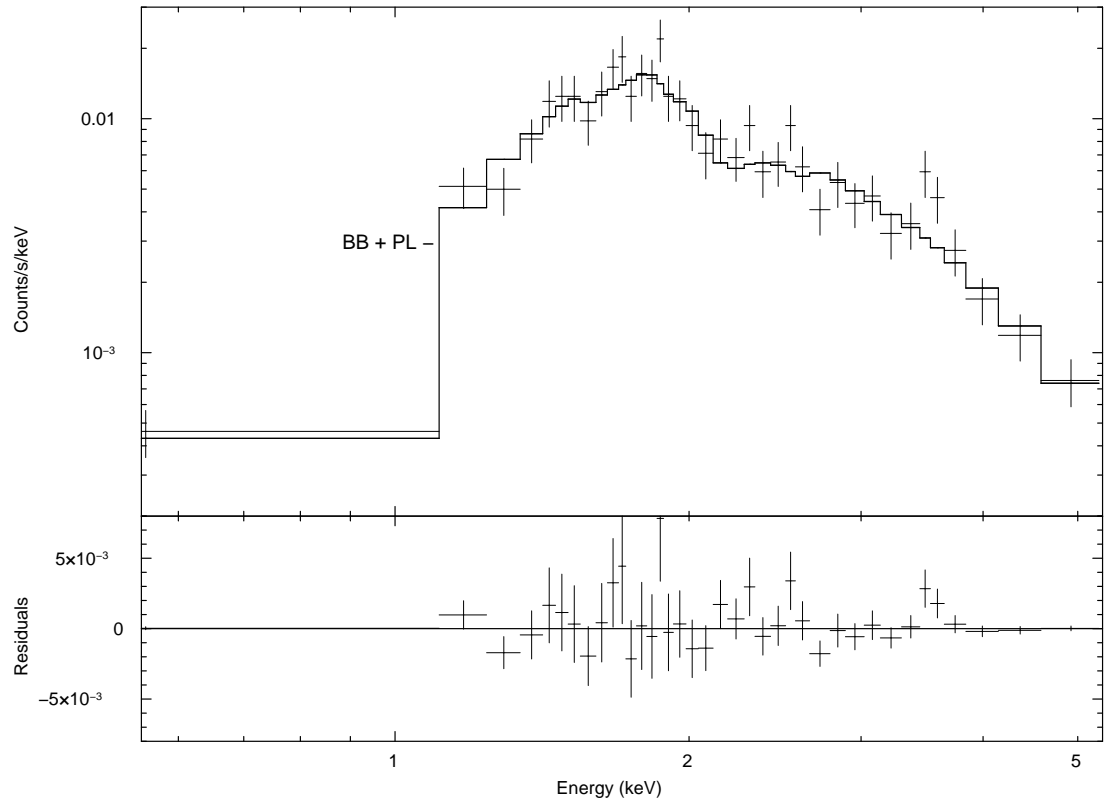}
\caption{
         Background-subtracted {\em Chandra} ACIS-S spectrum of the pulsar 
         CXO J164710.20$-$455217 obtained on 18-19 June 2005 (834 counts).
         The spectrum is rebinned to a minimum of 20 counts per bin.
         The overlaid model is a 2-component blackbody $+$ power-law
         (model C in Table 1).}
\end{figure}

\clearpage
\begin{figure}
\figurenum{2}
\epsscale{1.0}
\includegraphics*[width=12.5cm,angle=-90]{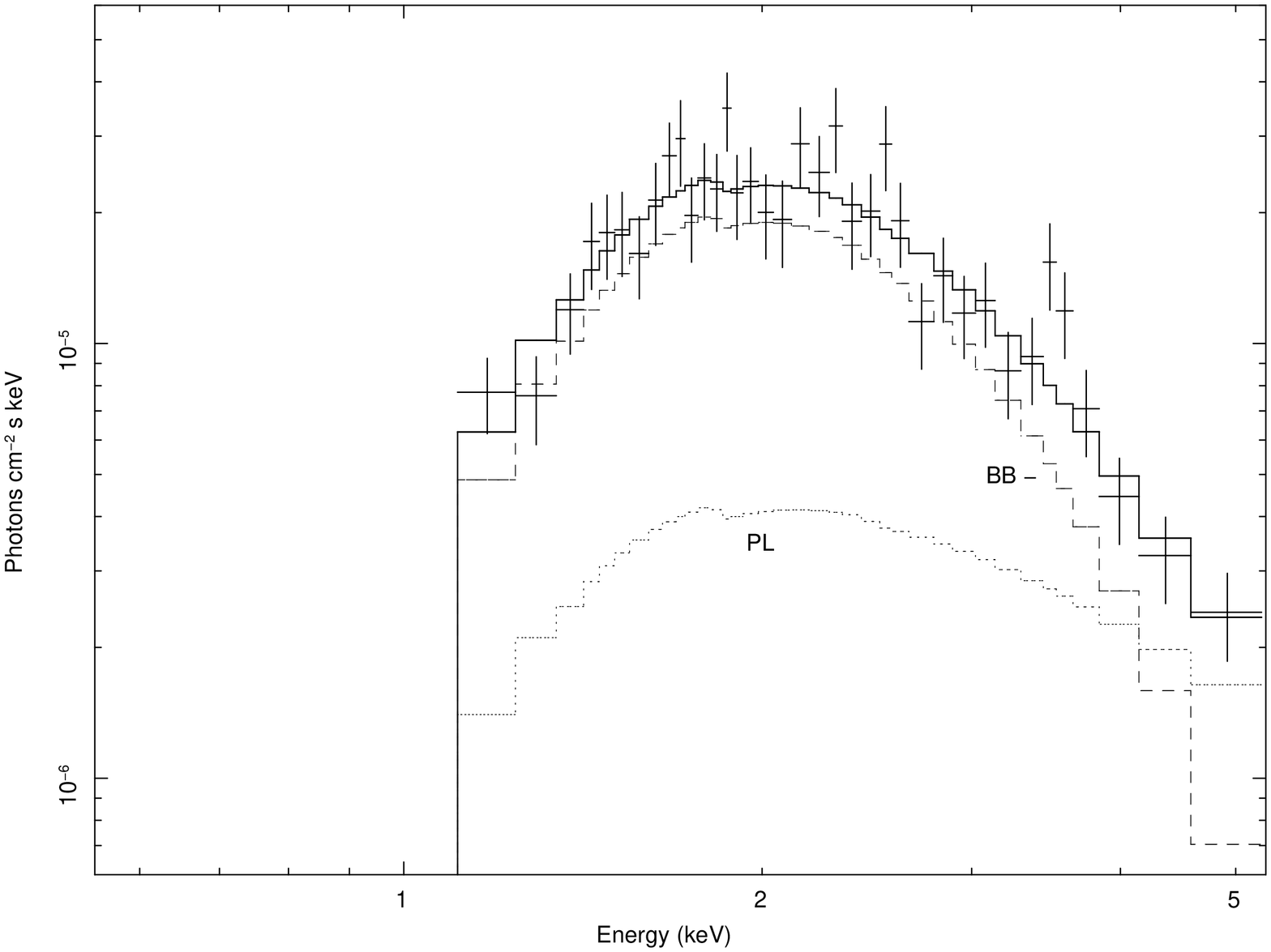}
\caption{
         Same as Figure 1 showing the unfolded spectrum.}
\end{figure}

\clearpage
\begin{figure}
\figurenum{3}
\epsscale{1.0}
\includegraphics*[width=12.5cm,angle=-90]{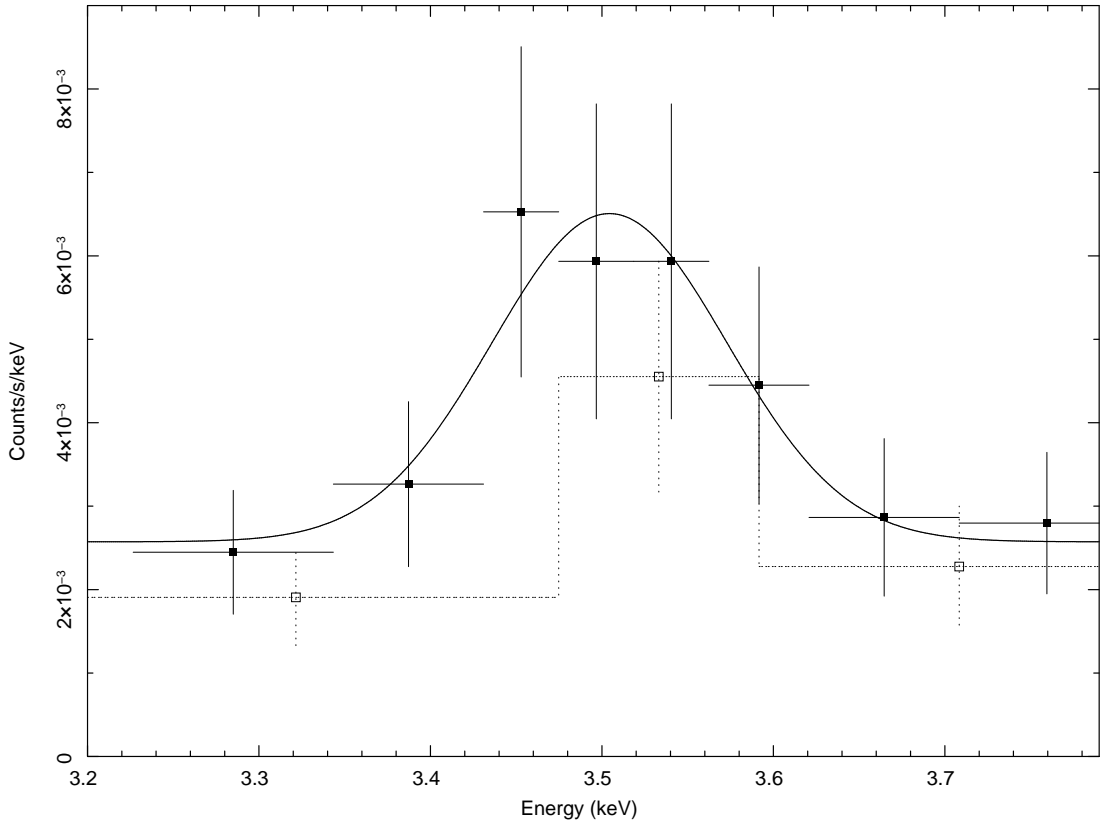}
\caption{
         {\em Chandra} ACIS-S spectra of the pulsar 
          CXO J164710.20$-$455217 showing the faint feature near 3.5 keV.
          The spectra are rebinned to a minimum of 10 counts per bin.
          Open squares are from the May 2005 observation (18.8 ksec) and 
          solid squares are from June 2005 (38.5 ksec). The Gaussian 
          fit to the second observation is centered at E = 3.504 keV.
          The feature contains 23 net counts [3.4 - 3.6 keV] above
          the continuum in the second observation and background is
          negligible ($<$1 count in this narrow energy range).}
\end{figure}

\clearpage
\begin{figure}
\figurenum{4}
\epsscale{1.0}
\includegraphics*[width=12.5cm,angle=-90]{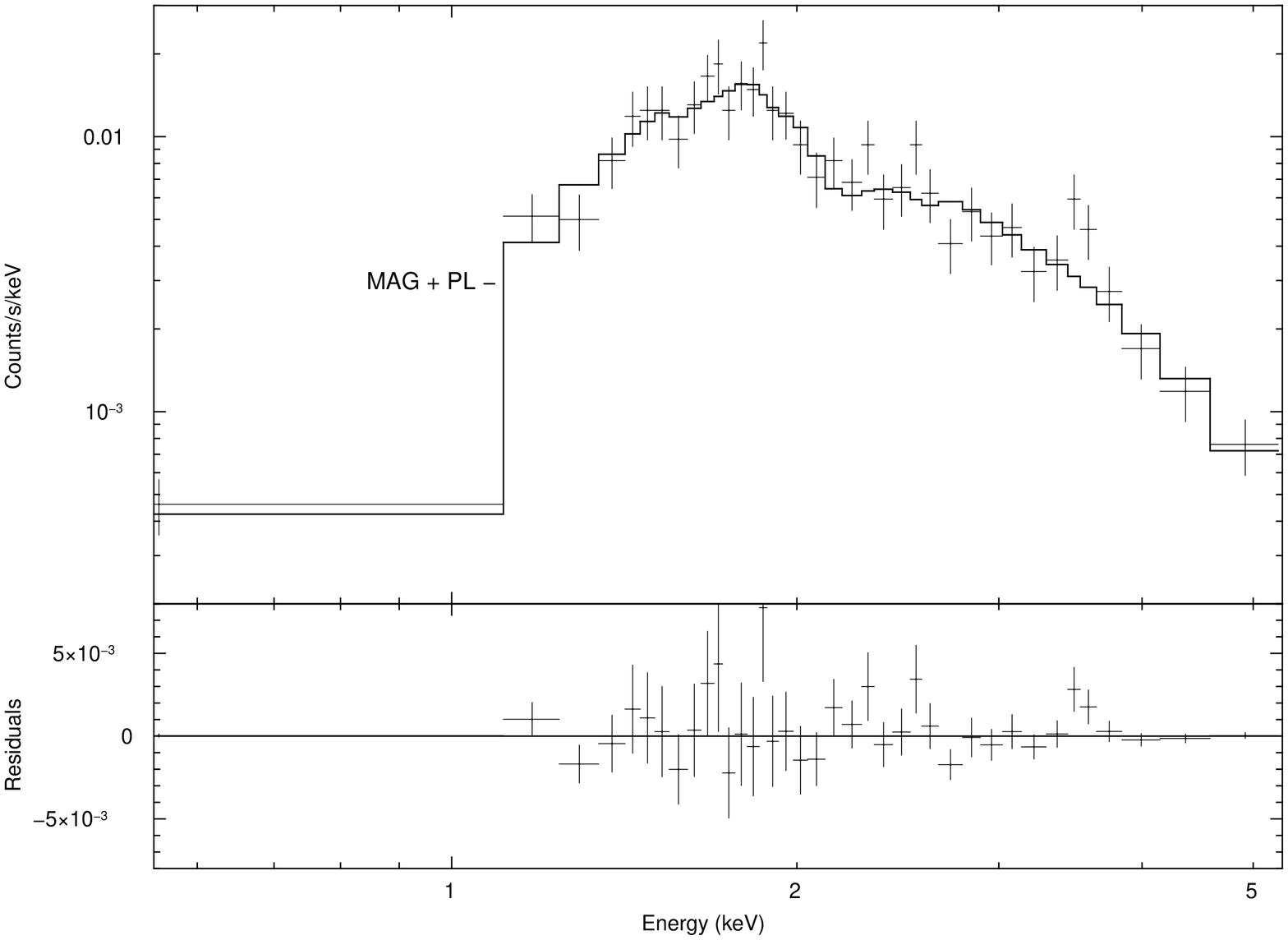}
\caption{
         Background-subtracted {\em Chandra} ACIS-S spectrum of the pulsar 
         CXO J164710.20$-$455217 obtained on 18-19 June 2005 (834 counts).
         The spectrum is rebinned to a minimum of 20 counts per bin.
         The overlaid model is a 2-component magnetar $+$ power-law
         (model E in Table 1).}
\end{figure}

\clearpage
\begin{figure}
\figurenum{5}
\epsscale{1.0}
\includegraphics*[width=12.5cm,angle=-90]{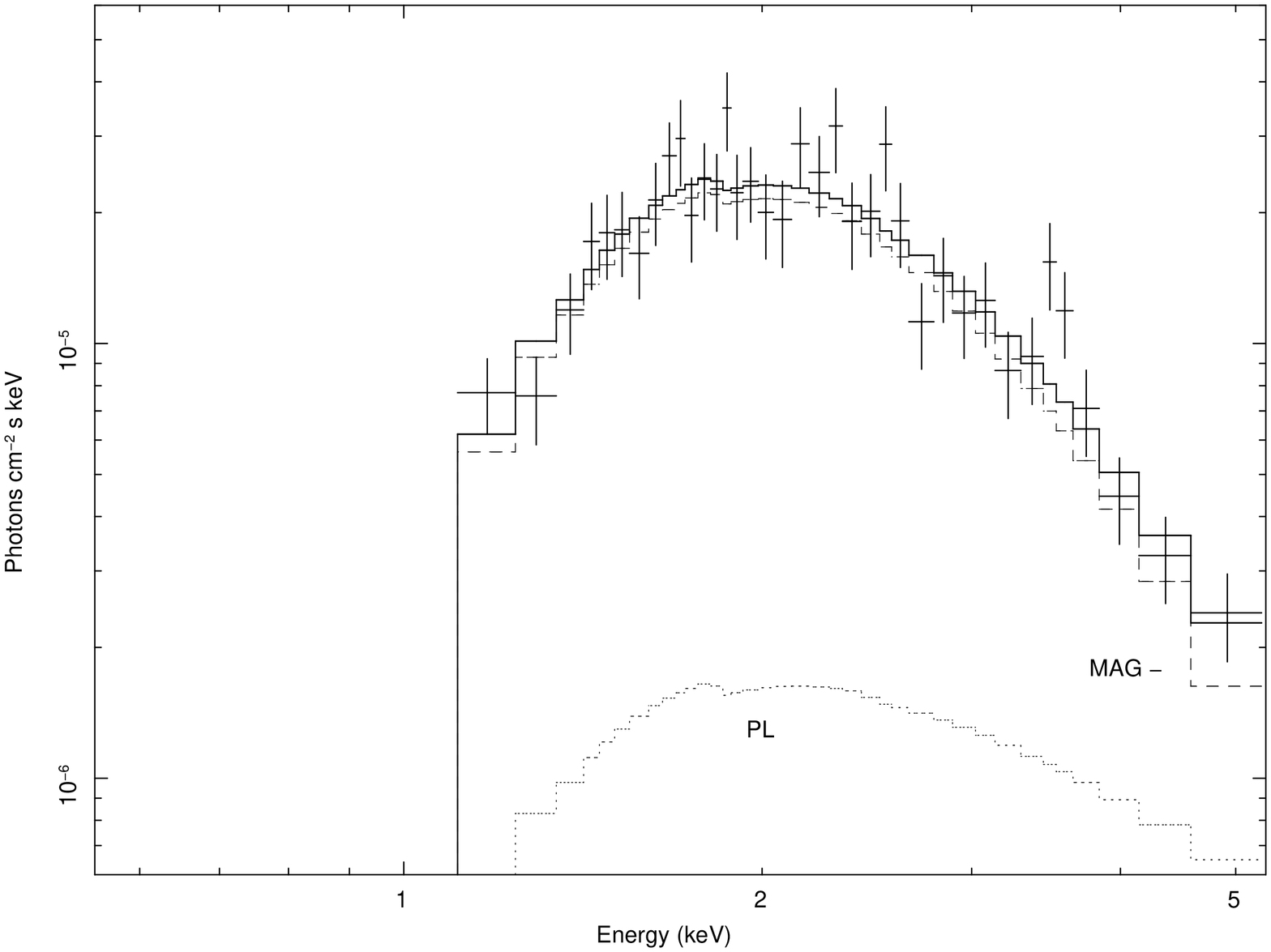}
\caption{
         Same as Figure 4 showing the unfolded spectrum.}
\end{figure}

\end{document}